\documentstyle[11pt,emulateapj]{article}

\begin{document}

\title{The Variability of Seyfert 1.8 and 1.9 Galaxies at 1.6 microns}

\author{A. C. Quillen\altaffilmark{1},
Shanna Shaked,
A.\ Alonso-Herrero,
Colleen McDonald,
Ariane Lee,
M.~J.\ Rieke, \&
G.~H.\ Rieke
}
\affil{Steward Observatory, The University of Arizona, Tucson, AZ 85721}
\altaffiltext{1}{aquillen@as.arizona.edu}

\begin{abstract}
We present a study of Seyfert 1.5-2.0 galaxies
observed at two epochs with the Hubble Space Telescope (HST) at 1.6 microns.
We find that unresolved nuclear emission from 9 of 14 
nuclei varies at the level of 10-40\% on timescales
of 0.7-14 months, depending upon the galaxy. 
A control sample of Seyfert galaxies lacking unresolved sources
and galaxies lacking Seyfert nuclei show
less than 3\% instrumental
variation in equivalent aperture measurements.
This proves that the unresolved
sources are non-stellar and associated with the central pc
of active galactic nuclei.
Unresolved sources in Seyfert 1.8 and 1.9 galaxies are not usually 
detected in HST optical surveys, however high angular resolution 
infrared observations will provide a way to measure time delays 
in these galaxies.

\end{abstract}

~ ~ ~
\section{Introduction}

The discovery that Seyfert 2 galaxies such as 
can have reflected or polarized broad line emission 
has led to an approach coined `unification' towards interpreting the
differences between active galactic nuclei (AGNs) in terms of orientation 
angle (\cite{antonucci}).  
The dusty torus of this unification paradigm
absorbs a significant fraction of the optical/UV/X-ray
luminosity of an active galaxy and consequently reradiates this energy at
infrared wavelengths NGC~1068.
As a result of this extinction it is difficult to observe
continuum radiation from Seyfert 2 galaxies at optical and
UV wavelengths (e.g., \cite{mulchaey}).   
An additional complication is that 
in a given aperture it may be difficult to identify
the percentage of flux from a non-stellar nuclear source 
(e.g., \cite{almu96}).
For example in Seyfert 2 galaxies much of the nuclear emission 
may originate from nuclear star formation 
(e.g., \cite{maiolino}; \cite{gonzalez}).

The high sensitivity and resolution of
near infrared imaging with the Hubble Space Telescope (HST) using 
NICMOS  (the Near Infrared Camera and Multi-Object Spectrograph)
allows us to probe galactic centers at wavelengths which experience
reduced extinction compared to the optical, and 
with a beam area about 30
times smaller than is typically achieved with ground-based observations at
these wavelengths.
This enables us to separate the nuclear emission from that of
the surrounding galaxy with unprecedented accuracy.
Though a previous survey using WFPC2 at $0.6\micron$ did
not detect unresolved nuclear continuum emission from Seyfert 2
galaxies (\cite{malkan}), about 60\% of the RSA and CfA samples 
(described below)
of Seyfert 1.8-2.0 galaxies display prominent unresolved nuclear sources
with diffraction rings in NICMOS images at $1.6\micron$ (\cite{colleen}).
%
Though we suspect that these unresolved continuum sources
are most likely associated directly with an AGN, they could
also be from unresolved star clusters, which are found in
a number of normal galaxies (\cite{carollo}).

Variability observed in the continuum (e.g., \cite{fitch})
is an intrinsic property of active galactic nuclei (AGNs) 
which demonstrates that the energy causing the emission
must arise from a very small volume.  This led early studies to 
suggest that accretion onto a massive black hole is responsible
for the luminosity  (\cite{salpeter}; \cite{zeldovich}).
Long term multi-year monitoring programs have found that Seyfert
1 galaxies are variable in the near-infrared 
(\cite{clavel3}; \cite{lebofsky}), however 
these programs have only seen a few Seyfert 2 nuclei
vary  (e.g., \cite{glass}; \cite{lebofsky}).  
Evidence for variability in the unresolved sources seen
in HST observations of Seyfert 2 galaxies
would provide evidence that this nuclear emission 
is non-stellar and so arises from the vicinity of a massive black hole.

\section{Observations}

In this paper we present a study of variability in Seyfert galaxies.
We have searched the HST archive for galaxies (Seyfert and normal)
which were imaged {\it twice} by HST at $1.6\micron$ in the F160W filter
with the NICMOS cameras.  
The Seyfert galaxies with unplanned duplicate observations 
either satisfy the Revised Shapely-Ames Catalog criterion
(described by Maiolino \& Rieke 1995) or are 
part of the CfA redshift survey (Huchra \& Burg 1992).
The Seyfert observations are discussed in 
Regan \& Mulchaey (1999) and \cite{martini} and the observations of
the normal or non-Seyfert galaxies are described by 
\cite{seigar_} and \cite{boker_}.
The observations are listed in Table 1 and are
grouped by the NICMOS cameras in which they were observed.
Images were reduced with the nicred data reduction
software (\cite{mcl}) using on orbit darks
and flats.  Each set of images in the F160W filter
was then combined according to the position observed.
The pixel sizes for the NICMOS cameras  are $\sim 0.043$,
$0.076$ and $0''.204$ for Cameras 1, 2 and 3 respectively.

At the center of these galaxies we expect 
contribution from an underlying stellar component 
in addition to that from an unresolved non-stellar component.
To measure the flux from the unresolved component
we must subtract a resolved stellar component.
However this procedure is dependent upon
assumptions made about the point spread function,
the form of the stellar surface brightness profile fit to the image
and the region over which we fit this profile.
This procedure adds uncertainty in the measurement
of the unresolved component.  
However, aperture photometry has proved quite robust
with observations of flux calibration standard stars
showing variation less than 1\% over the lifetime of NICMOS 
(M.~Rieke, private communication 1999).
We therefore opt to use aperture photometry to measure flux
variations, and then subsequently correct for contamination
of the aperture by the background galaxy.   

From each pair of images we measure fluxes in apertures of the 
same angular size.  No background was
subtracted since the level of background expected at $1.6\micron$
is negligible compared to the galaxy surface brightnesses.    
Apertures are listed in Table 1
and were chosen so that more than 75\% of the flux of an unresolved
source would be contained in the aperture.  
We chose apertures based on which two cameras
were used to observe the object.
We list in Table 1 the difference divided by the mean of the 
two flux measurements for each pair of images.

To determine whether the nuclear sources are variable we 
need to quantify the 
level of intrinsic scatter in our flux measurements.
As a control sample we use the galaxies not identified
as Seyfert galaxies and those containing Seyfert nuclei but
lacking an unresolved nuclear component.
Comparing Camera 2 and Camera 3 measurements for this control sample 
we find a mean difference of 
$\mu=-0.9 \pm 0.7\%$ with a variance of $\sigma =2.0\%$
in the measurements.  
Comparing measurements with two observations in Camera 2 
for this control sample 
we find a mean difference of
$\mu=-0.6 \pm 0.6\%$ with a variance of $\sigma =1.4\%$.
Unfortunately our control sample only contains 2 galaxies with 
observations in Camera 1 and Camera 2 (MRK~266 and NGC~5929).
To supplement this we also measured stars observed in both Camera 1 and 2
in the vicinity of the Galactic Center. 
Differences in fluxes measured in these 3 image pairs were less than 3\%. 
The statistics of our control sample suggest that the intrinsic
scatter of our measurements is smaller than a difference of 3\%
for all pairs of images.
We therefore estimate that flux differences greater than 6\%
are statistically significant (at $\gtrsim 2 \sigma$ level) 
and likely to be caused by variability
and not by scatter in the measurements.
The galaxies in which we measure differences larger than this level
are listed in Table 2.

We did not find that the unresolved nuclear sources in NGC~404 
our NGC~2903 were variable.
As demonstrated with UV spectra by \cite{maoz_}, 
it is possible that the unresolved component in NGC~404 is from 
a young star cluster.
The same is probably true in NGC~2903 which also contains
a compact nuclear source and has a nuclear HII region type spectrum.
The scatter in our aperture measurements does not appear to be 
dependent on the surface brightness profile of the galaxy.
No large differences were measured between image pairs for
galaxies lacking an unresolved nuclear source.

To estimate the level of variability in the unresolved component
we must measure the contribution within the aperture of this component.
For each camera we measured a point spread function from stars
in the images.  We then fit the sum of an exponential
bulge profile and the point spread function to the surface brightness
profile.  The error in this procedure we estimated from the scatter
in the residuals and was about $\pm 15\%$ of the total 
unresolved flux measured.
We used the flux from the unresolved component and the shape of the 
point spread function to estimate the contribution to the flux measured 
in the apertures listed in Table 1.   
The differences in the aperture flux measurements are lower limits 
for differences in the fluxes of the unresolved components
(in the limit that the galaxy contributes no flux in these apertures).
The mean unresolved fluxes from the two epochs
and extent of variability of the unresolved components 
(the difference divided by the mean) are listed in Table 2.

\section{Discussion}

In 9 out of 14 Seyfert 1.5-2.0 galaxies with unresolved
components we find
a variation greater than 10\% in the flux of their unresolved
continuum nuclear sources in 2 epochs of observations at $1.6\micron$.
A control sample of Seyfert galaxies lacking unresolved sources
and galaxies lacking Seyfert nuclei show 
less than 3\% instrumental
variation in equivalent aperture measurements.
This suggests that the variability detected is statistically significant
at the level of $\gtrsim 2 \sigma$.
Since we see variations between 0.7-14 month timescales
the unresolved sources are probably non-stellar and associated
with the central pc of active galactic nuclei.
The luminous Seyfert 1 galaxy in our sample, NGC~4151, shows a variation
of 10\% in its nuclear flux, similar to that seen in the 
other Seyfert galaxies.

From Table 2 we see that most of the variable sources are Seyfert 1.8
or 1.9 galaxies.  
NGC~1275, NGC~5033, and NGC~5273 are usually classified as Seyfert 
1.9 galaxies though \cite{hob} classify them as S1.5.
There are two Seyfert 2 galaxies exhibiting variability: MRK~533
and NGC~5347.  In MRK~533 a broad component in Pa$\alpha$ was detected by
\cite{ruiz_} and so this galaxy could be classified as a Seyfert 1.9.
Seyfert 1.8 and 1.9 galaxies are more likely to display
unresolved nuclear sources than Seyfert 2.0 galaxies (\cite{colleen}).
In the context of the unification model, 
reduced extinction towards the continuum emitting region at
$1.6 \micron$ would be expected 
in Seyfert galaxies which display faint broad line emission.
However, this might also suggest that the sizes of the Broad Line Region 
and 1.6 micron
continuum emission region are small compared to the material
responsible for the bulk of the extinction.

Two major sources for AGN continuum variability are generally
considered:  1) instabilities in an accretion disk
(e.g., \cite{shakura}) and
2) jet related processes (e.g., as discussed by 
Tsvetanov et al.~1998).  
The second case could be a possible explanation for variability
NGC~1275 since it is bright at radio wavelengths and is significantly
polarized at optical wavelengths (as discussed by \cite{angel}).
However, the luminosity of the compact nucleus of this galaxy
at 1.3 GHz is about 20 times lower than that we measure at 1.6 microns  
(using the flux from \cite{taylor}).
So the $1.6\micron$ flux is higher than what would be expected from
synchrotron emission and could be from an additional thermal
component (e.g., from hot dust).
Better measurements showing the shape of the spectral energy distribution 
spanning the optical and near-infrared region (to see if two components
are present) or a polarization measurement at $1.6\micron$ would help
differentiate between a thermal or non-thermal origin for
the near-infrared emission.

For the remainder of the Seyferts, their low radio power implies that
jet related processes are not responsible for the variability.
From observations of the Seyfert 1 galaxy
Fairall 9, \cite{clavel_} observed
large, 400 day, time delays between variations seen
at 2 and $3\micron$ and those seen in the UV.
Little or no time delay was seen
at $1.2\micron$.  This led them to suggest
that the longer wavelength emission was associated with
hot dust located outside the Broad Line Region 
(e.g., \cite{lebofsky}; \cite{barvainis}; \cite{laor})
and that the shorter
wavelength emission was reprocessed near the UV emitting region.

For hot dust to cause the $1.6\micron$ emission,
dust grain temperatures resulting from absorption of UV radiation
must be quite high, nearly that expected for 
sublimation ($T \sim 2000$K).
The grain temperature should reach this level at a radius
$r \sim 0.06 {\rm pc} \left({L \over 10^{44} {\rm erg/s}}\right)^{1/2}$ 
(following the estimate given in \cite{barvainis}).
This radius would have a characteristic variability timescale
of $\sim 70$ days or 2 months for a source of $10^{44}$ ergs/s.
We can crudely estimate the bolometric luminosity of our sources
from that at $1.6\micron$ (which are listed in Table 2) 
by assuming a ratio of $\sim 10$ between
the $1.6\micron$ and mid-IR luminosity (e.g, \cite{fadda}
for the Seyfert 2s) and
a ratio of $\sim 10$ between the mid-IR and bolometric luminosity
(e.g., \cite{spinoglio}).  The timescales over which
we see variations for the brighter sources such as NGC~1275, MRK~533
and UGC~12138 ($L \sim 10^{44}$ ergs/s)  
are consistent with the 2 month minimum estimated
for emission from hot dust.
The least luminous of our sources, NGC~4395 ($L \sim 10^{41}$ ergs/s),
could have a variability timescale 
of only a few days for hot dust emitting at $1.6 \micron$, 
again consistent with the timescale (a few weeks) 
over which we see a variation.

Emission from hot dust may not necessarily
dominate at $1.6\micron$ 
since the emitting material would require a
temperature near the sublimation point
of graphites and silicates (\cite{laor}). 
However transient super heating at larger radii 
could still cause emission from hot dust at this wavelength.  
While the timescales over which we see variability are 
comparable to those expected from hot dust near a sublimation radius,
a long term study comparing
flux variations between the near-infrared and X-ray emission
would be needed to determine the exact nature of the $1.6\micron$ emission.
This kind of study 
would also place strong constraints on disk and torus models 
for the infrared emission (e.g., \cite{efstathiou}; \cite{fadda}).
 
Most of the unresolved nuclear sources studied
here exhibit variability. This suggests that most
of the many unresolved continuum sources recently discovered 
in near-infrared surveys (\cite{colleen}; \cite{almu96})
(and not seen in previous optical surveys) are 
non-stellar and associated with the central pc of an AGN.  
The near-infrared continuum in low luminosity AGNs can now
be studied in a set of objects comprising a larger
range of luminosity and orientations.  This should provide tests of
the unification model for Seyfert 1 and 2 galaxies as well
as the nature of accretion in these lower luminosity sources.

\acknowledgments

We thank the referee, Ski Antonnuci, for comments which 
have improved this paper.
We thank Brad Peterson and Chien Peng for helpfull discussions on this work.
Support for this work was provided by NASA through grant number
GO-07869.01-96A
from the Space Telescope Science Institute, which is operated 
by the Association
of Universities for Research in Astronomy, Incorporated, under NASA
contract NAS5-26555.
We also acknowledge support from NASA project NAG-53359 and
NAG-53042 and from JPL Contract No.~961633.


{ }

\begin{table}
\caption{Multi Epoch Aperture Photometry of Seyfert and Normal Galaxies at $1.6\micron$
\label{tbl-1}}
\begin{tabular}{lccccccccc}
\tableline
\tableline
Galaxy & Type & Nuc. & Prop1 & Date1 & Prop2 & Date2 & Flux1 & Flux2 & Diff\%  \\
(1)    & (2)  &  (3) &  (4)  &  (5)  &  (6) &  (7)   &   (8) &   (9) &  (10)\\
\tableline
IC  5063& S2$^d$          &D& 7330/2& 25/09/97& 7119/2& 18/04/97& 2.74& 2.73&  0.57   \\
NGC~1275&S1.9$^a$/S1.5$^b$&D& 7330/2& 16/03/98& 7457/2& 15/08/97& 3.74& 3.10& 19.0    \\
NGC~2460&                 &R& 7330/2& 28/02/98& 7331/2& 11/09/97& 3.09& 3.06&  1.06   \\ 
NGC~2985& T1.9$^b$        &R& 7330/2& 18/05/98& 7331/2& 13/09/97& 5.44& 5.44& -0.04   \\ 
NGC~3368&  L2$^b$         &R& 7330/2& 08/05/98& 7331/2& 04/05/98& 9.23& 9.33& -1.10   \\
NGC~2903&  H$^b$          &W& 7330/2& 22/04/98& 7331/2& 02/10/97& 2.96& 3.01& -1.66   \\
NGC~6951&  S2$^b$         &R& 7330/2& 30/03/98& 7331/2& 16/12/97& 3.88& 3.86&  0.53   \\
NGC~7177&  T2$^b$         &R& 7330/2& 15/06/97& 7331/2& 19/09/97& 2.40& 2.46& -2.63   \\
\tableline
MRK~266 & S2$^a$          &R& 7867/1& 30/04/98& 7328/2& 13/09/97& 1.10& 1.07&  2.32   \\
MRK~573 & S2$^a$          &D& 7867/1& 27/08/98& 7330/2& 26/06/97& 2.26& 2.20&  2.69   \\
NGC~3982& S2$^b$          &F& 7867/1& 10/09/98& 7330/2& 22/06/97& 1.36& 1.34&  1.91   \\
NGC~5033&S1.9$^a$/S1.5$^b$&D& 7867/1& 28/04/98& 7330/2& 19/08/97& 5.23& 6.31& -8.78   \\
NGC~5252& S1.9$^a$        &D& 7867/1& 09/03/98& 7330/2& 05/04/98& 1.81& 1.80&  0.24   \\
NGC~5273&S1.9$^a$/S1.5$^b$&D& 7867/1& 27/05/98& 7330/2& 03/04/98& 2.35& 2.57& -8.63   \\
NGC~5347& S2$^e$          &D& 7867/1& 06/11/98& 7330/2& 02/09/97& 2.30& 2.18&  5.45   \\
NGC~5929& S2$^a$          &R& 7867/1& 17/07/98& 7330/2& 21/05/98& 1.91& 1.90&  0.43   \\
\tableline
MRK~471 & S1.8$^c$        &D& 7328/1& 16/04/98& 7867/1& 08/07/97& 0.88& 0.86&  2.06    \\
MRK~533 & S2$^a$          &D& 7328/1& 13/09/98& 7867/1& 05/09/97& 4.86& 5.34& -9.31    \\
UGC 12138&S1.8$^a$        &D& 7328/1& 09/09/98& 7867/1& 28/07/97& 3.60& 4.23& -16.0    \\
UM  146 & S1.9$^a$        &D& 7328/1& 03/08/98& 7867/1& 13/09/97& 1.36& 1.46& -6.89    \\
\tableline
NGC~1241& S2$^f$          &W& 7330/2& 18/03/98& 7919/3& 19/06/98& 4.24 & 4.24 &  0.01   \\
NGC~214 &                 &R& 7330/2& 29/05/98& 7919/3& 09/06/98& 2.73 & 2.67 &  2.13   \\
NGC~2639& S1.9$^b$        &R& 7330/2& 23/02/98& 7919/3& 07/06/98& 7.25 & 7.27 & -0.29   \\
NGC~2903& H$^b$           &W& 7330/2& 22/04/98& 7919/3& 09/06/98& 5.31 & 5.28 &  0.45   \\
NGC~3627& T2$^b$          &R& 7330/2& 22/04/98& 7919/3& 04/06/98& 14.99& 15.59& -3.89   \\
\tableline
\end{tabular}
\end{table}

\begin{table}
\begin{tabular}{lccccccccc}
\multicolumn{10}{l}{Table 1 continued} \\
\tableline
NGC~404 & L2$^b$          &D& 7330/2& 02/03/98& 7919/3& 19/01/98& 11.74& 11.64&  0.84   \\
NGC~4151& S1.5$^b$        &D& 7215/2& 22/05/98& 7806/3& 14/10/97& 97.35&108.99& -11.3   \\
NGC~4258& S1.9$^b$        &W& 7330/2& 21/11/97& 7919/3& 09/06/98& 19.20& 18.84&  1.86   \\
NGC~4395& S1.8$^b$        &D& 7330/2& 17/05/98& 7919/3& 07/06/98& 1.14 & 1.02 & 11.2    \\
NGC~5128& S2$^g$          &D& 7330/2& 17/09/97& 7919/3& 17/06/98& 16.27& 16.72& -2.72   \\
NGC~628 &                 &W& 7330/2& 15/06/97& 7919/3& 30/01/98& 1.37 & 1.40 & -2.42   \\
NGC~6744& L$^h$           &R& 7330/2& 09/09/97& 7919/3& 11/06/98& 7.03 & 7.18 & -2.08   \\
NGC~6946& H$^b$           &R& 7330/2& 18/05/98& 7919/3& 19/01/98& 13.73& 13.99& -1.86   \\
\tableline
\tablecomments{ \baselineskip=12.5pt 
Seyfert and normal galaxies have been grouped by the NICMOS cameras in which they 
were observed.
The first group consists of Camera 2/Camera 2 pairs, the second Camera 1/Camera 2 pairs,
the third Camera 1/Camera 1 pairs and the last group Camera 2/Camera 3 pairs.
Columns:  (1) Galaxy; 
(2) Classification of emission lines in the nucleus. 
References are denoted with superscripts:
$^a$=Osterbrock \& Martel (1993),
$^b$=Ho, Filippenko \& Sargent~(1995a,b) (classifications from these works
include H = HII nucleus, S = Seyfert nucleus, L = LINER and T = transition
object with numbers corresponding to subtypes),
$^c$=Dahari, \& De Robertis (1988),
$^d$=A polarized broad line component was detected in IC~5063 by Inglis et al.~(1993),
$^e$=Huchra \& Burg (1992). No data are available about the line ratios of
NGC~5347,
$^f$=Dahari (1985),
$^g$=Tadhunter et al.~(1993),
Spectroscopic identifications for the nuclei of NGC~2460 and NGC~214 could
not be found.  The nucleus of NGC~628 lacks emission lines (Ho et al.~1995a),
$^h$= NGC~6744 was classified as a LINER by 
Vaceli et al.~(1997) and no subtype was given; 
(3) Type of nucleus seen in the F160W images.  When the nucleus
displayed a clear diffraction ring we denote `D', when the ring
was faint we denote
`F', and when the galaxy was resolved we denote `R'.  When there
was an unresolved peak but no sign of a diffraction ring we denote `W';
(4) Proposal ID number followed by camera number of the first NICMOS image considered;
(5) Date that this image was observed;
(6) Proposal ID number followed by camera number of the second NICMOS image considered;
(7) Date that this image was observed;
(8) Nuclear flux at $1.6\micron$ 
measured in mJy for the image identified by columns 4 and 5.
For the galaxies observed in Camera 1 and 2 we used an aperture of $0''.602$ in diameter.
For the galaxies observed solely in Camera 1 the aperture was $0''.602$.
For the galaxies observed solely in Camera 2 the aperture was $0''.760$.
For the galaxies observed in Camera 2 and 3 the aperture was $1''.216$;
To convert these fluxes into mJy we used  conversion factors
$2.360\times10^{-3}$, $2.190\times10^{-3}$, $2.776\times10^{-3}$ 
mJy per DN/s for Cameras 1, 2 and 3 respectively.
This flux calibration is based on
measurements of the standard stars P330-E and P172-D during
the Servicing Mission Observatory Verification program 
and subsequent observations (M.~Rieke 1999, private communication);
(9) Flux in an aperture for the image identified by columns 6 and 7;
(10) Percent difference divided by the mean of the fluxes listed in columns 7 and 8.
}
\end{tabular}
\end{table}

\begin{deluxetable}{lcccccccc}
\footnotesize
\tablecaption{Variable Unresolved Nuclear Sources \label{tbl-2}}
\tablehead{
\multicolumn{1}{l}{Galaxy} &
\colhead{Type}             &
\colhead{$v_{\rm helio}$}  &
\colhead{size}             &
\colhead{flux}             &
\colhead{var}              &  
\colhead{time}             &
\colhead{$L_{1.6 \micron }$} \\  
\colhead{}           &   
\colhead{}           &   
\colhead{km/s}       &   
\colhead{pc}         &   
\colhead{mJy}        &   
\colhead{\%}         &   
\colhead{months}     & 
\colhead{ergs/s}     \\  
\colhead{(1)}        &   
\colhead{(2)}        &   
\colhead{(3)}        &   
\colhead{(4)}        &   
\colhead{(5)}        &  
\colhead{(6)}        &          
\colhead{(7)}        &  
\colhead{(8)}            
}
\startdata
NGC~1275$^a$ &S1.9/S1.5&5264& 34     & 3.20 & 20  & 7.0  & 3.3e42 \\
NGC~5033     &S1.9/S1.5&875 & 6      & 2.41 & 45  & 8.3  & 7.0e40 \\
NGC~5273     &S1.9/S1.5&1089& 7      & 1.22 & 18  & 1.8  & 5.4e40 \\
MRK~533      &S2$^b$&  8713 & 56     & 3.59 & 13  & 14.1 & 1.0e43 \\
NGC~5347     &  S2  &  2335 & 15     & 1.08 & 11  & 12.3 & 2.2e41 \\
UGC~12138    &  S1.8&  7375 & 48     & 2.25 & 28  & 13.4 & 4.6e42 \\
UM~146       &  S1.9&  5208 & 33     & 0.60 & 17  & 10.7 & 6.1e41 \\  
NGC~4395$^c$ &  S1.8&  319  & 1.3    & 0.83 & 15  & 0.7  & 1.2e39 \\
NGC~4151     &  S1.5&  995  & 6.5    & 103  & 11  & 7.3  & 3.8e42 \\
\enddata
\tablenotetext{a}{Variation in the nuclear flux of NGC~1275 was reported 
previously by Lebofsky and Rieke (1980).}
\tablenotetext{b}{In MRK~533 a broad component in Pa$\alpha$ was detected by
Ruiz, Rieke \& Schmidt (1994).  
}
\tablenotetext{c}{NGC~4395 has been labelled `the least luminous Seyfert 
1 galaxy' (Filippenko \& Sargent 1989; Filippenko, Ho, \& Sargent 1993). 
The flux at $4400 \AA$ has varied by a factor of three in just one month
(Lira et al.~1999), so a variation of 15\% in a few weeks at $1.6\micron$
is not surprising.}
\tablecomments{
The Seyfert galaxies with unresolved sources which did not 
vary significantly between observations 
were IC~5063, MRK~573, NGC~5252, MRK~471 and NGC~5128.
Columns:- (1) Seyfert Galaxy; (2) Seyfert type; 
(3) Heliocentric velocity;
(4) Physical size corresponding to $0''.1$. These have
been estimated using a Hubble constant of 75 km s$^{-1}$ Mpc$^{-1}$
except in the case of NGC~4395 for which we adopt a distance of 2.6 Mpc
(Rowan-Robinson 1985);
(5) The flux of the unresolved component (galaxy subtracted)
averaged between the two measurements.
We estimate the error to be $\sim \pm 15\%$ of the flux listed;
(6) Percent variation (absolute value of the difference 
divided by the mean) of the unresolved component;
(7) Time between the two different observations;
(8) Mean luminosity at $1.6\micron$ estimated by $\nu f_\nu$.
}
\end{deluxetable}


\begin{thebibliography}{ }


\bibitem[Alonso-Herrero et al.~1996]{almu96}
Alonso-Herrero, A., Ward, M.~J., \& Kotilainen, J.~K.\ 1996, MNRAS, 273, 902

\bibitem[Angel \& Stockman 1980]{angel}
Angel, J.~R.~P., \& Stockman, H.~S.\ 1980, ARA\&A, 18, 321

\bibitem[Antonucci 1993]{antonucci}
Antonucci, R.~R.~J.\ 1993, ARA\&A, 31, 473


\bibitem[Barvainis 1987]{barvainis}
Barvainis, R.\ 1987, ApJ, 320, 537



\bibitem[B\"oker et al.~(1999)]{boker_}
B\"oker, T., Calzetti, D., Sparks, W., Axon, D., Bergeron, L. E., 
Bushouse, H., Colina, L., Daou, D., Gilmore, D.,
Holfeltz, S., Mackenty, J., Mazzuca, L., Monroe, B., Najita, J.,
Noll, K., Nota, A., Ritchie, C., Schultz, A., Sosey, M., Storrs, A.,
\& Suchkov, A.~1999, ApJS, 124, 95



\bibitem[Carollo et al.~1997]{carollo}
Carollo, C.~M., Stiavelli, M., de Zeeuw, P.~T., \& Mack, J.~1998, 
AJ, 114, 2366


\bibitem[Clavel et al.~(1989)]{clavel_}
Clavel, J., Wamsteker, W., \& Glass, I.~S.\  1989, ApJ, 337, 236

\bibitem[Dahari \& De Robertis (1988)]{dahari}
Dahari, O., \& De Robertis, M.~M.~1988, ApJS, 67, 249

\bibitem[Dahari 1985]{dahari85}
Dahari, O.~1985, ApJS, 57, 643

\bibitem[Efstathiou \& Rowan-Robinson 1995]{efstathiou}
Efstathiou, A., \& Rowan-Robinson, M.\
1995, MNRAS, 273, 649

\bibitem[Fadda et al.~1998]{fadda}
Fadda, D., Giuricin, G., Granato, G.~L., \& Vecchies, D.\
1998, ApJ, 496, 117


\bibitem[Filippenko, Ho, \& Sargent(1993)]{filippenko93}
Filippenko, A.~V., Ho, L.~C., \& Sargent, W.~L.~W.\ 1993, ApJ, 410, L75

\bibitem[Filippenko \& Sargent(1989)]{filippenko89} 
Filippenko, A.~V., \& Sargent, W.~L.~W.\ 1989, ApJ, 342, L11

\bibitem[Fitch, Pacholczyk, \& Weymann 1967]{fitch}
Fitch, W., Pacholczyk, A.~G., \& Weymann, R.~J.~1967, ApJ, 150, L67 

\bibitem[Glass 1997]{glass}
Glass, I.~S.~1997, MNRAS, 292, L50

\bibitem[Gonzalez-Delgado \& Perez 1993]{gonzalez}
Gonzalez-Delgado, R.~M., \& Perez, E.~1993, Ap\&SS, 205, 127



\bibitem[Ho, Filippenko \& Sargent (1995)]{hoa}
Ho, L.~C., Filippenko, A.~V., \& Sargent, W.~L.~W.~1995a, ApJS, 98, 477

\bibitem[Ho, Filippenko \& Sargent (1995)]{hob}
Ho, L.~C., Filippenko, A.~V., \& Sargent, W.~L.~W.~1995b, ApJS, 112, 315

\bibitem[Huchra \& Burg]{huchra}
Huchra, J., \& Burg, R.~1992, ApJ, 393, 90

\bibitem[Inglis et al.~(1993)]{inglis}
Inglis, M.~D., Brindle, C., Hough, J.~H.,
Young, S., Axon, D.~J., Bailey, J.~A., \&
Ward, M.~J.~1993, MNRAS, 263, 895

\bibitem[Lebofsky \& Rieke 1980]{lebofsky}
Lebofsky, M.~J., \& Rieke G.~H.\ 1980, Nature, 284, 410

\bibitem[Lira et al.(1999)]{lira}
Lira, P., Lawrence, A., O'Brien, P., Johnson, R.~A., Terlevich, R.,  \&
Bannister, N., 1999, MNRAS, 305, 109 

\bibitem[Maiolino \& Rieke(1995)]{maiolino+r}
Maiolino, R., \& Rieke, G.~H.\ 1995, ApJ, 454, 95

\bibitem[Maiolino et al.~1997]{maiolino}
Maiolino, R., Ruiz, M., Rieke, G.~H., \& Papadopoulos, P.\
1997, ApJ, 485, 552


\bibitem[Malkan et al.~1998]{malkan}
Malkan, M.~A., Gorjian, V., \& Tam, R.\ 1998, ApJS, 117, 25

\bibitem[Maoz et al.~(1998)]{maoz_}
Maoz, D., Koratkar, A., Shields, J.~C., Ho, L.~C.,
Filippenko, A.~V., \& Sternberg, A.~1998, AJ, 116, 55

\bibitem[Martini \& Pogge~(1999)]{martini}
Martini, P.,  \& Pogge, R.~W.\ 1999,
Accepted for publication in AJ, (astro-ph/9909032)


\bibitem[McDonald et al.~2000]{colleen}
McDonald, C., Quillen, A.~C., Alonso-Herrero, A., Shaked, S., Lee, A., 
Rieke, M.~J., \& Rieke G.~H.\ 2000, in preparation 


\bibitem[Ruiz, Rieke, \& Schmidt~(1994)]{ruiz_}
Ruiz, M., Rieke, G.~H.,\& Schmidt, G.~D.\ 1994, ApJ, 423, 608

\bibitem[McLeod 1997]{mcl}
McLeod, B.~1997, proceedings of the 1997 HST Calibration Workshop,
eds.~S.~Casertano, R.~Jedrzejewski, T.~Keyes, and M.~Stevens,
published by the Space Telescope Science Institute, Baltimore, MD, p.~281

\bibitem[Mulchaey et al.~1994]{mulchaey}
Mulchaey, J.~S., Koratkar, A., Ward, M.~J., Wilson, A.~S.,
Whittle, M., Antonucci, R.~R.~J., Kinney, A.~L., \& Hurt, T.\
1994, ApJ, 436, 586

\bibitem[Netzer \& Laor 1993]{laor}
Netzer, H., \& Laor, A.\ 1993, ApJ, 404, L51


\bibitem[Regan \& Mulchaey(1999)]{regan}
Regan, M.~W., \& Mulchaey, J.~S.~1999, AJ, 117, 2676

\bibitem[Rowan-Robinson(1985)]{rowan}
Rowan-Robinson, M.\ 1985, The Cosmological Distance Ladder,
(New York: Freeman), page 171 

\bibitem[Salpter 1964]{salpeter}
Salpeter, E.~E.\ 1964, ApJ, 140, 796

\bibitem[Seigar et al.~(2000)]{seigar_}
Seigar, M., Stiavelli, M., Carollo, C.~M., de Zeeuw, P.~T., \& DeJongue, H.\ 2000,
`Galaxy Dynamics: from the Early Universe to the Present, 
15th IAP meeting held in Paris, France, July 9-13, 1999, 
Eds.~F.~Combes, G.~A.~Mamon, \&  V.~Charmandaris, 
to be published in the ASP Conference Series, 1999, page 115

\bibitem[Shakura \& Sunyaev 1973]{shakura}
Shakura, N.~I., \& Sunyaev, R.~A.~1973, A\&A, 24, 337

\bibitem[Spinoglio et al.~1995]{spinoglio}
Spinoglio, L., Malkan, M.~A., Rush, B., Carrasco, L., \& Recillas-Cruz, E.\
1995, ApJ, 453, 616

\bibitem[Tadhunter et al.~1993]{tadhunter}
Tadhunter, C.~N., Morganti, R., di Serego-Alighieri, S., Fosbury, R.~A.~E., 
Danziger, I.~J.~1993, MNRAS, 263, 999

\bibitem[Taylor \& Bermeulen]{taylor}
Taylor, G.~B., \& Vermeulen, R.~C.~1996, ApJ, 457, L69

\bibitem[Tsvetanov et al.(1998)]{tsvetanov}
Tsvetanov, Z.~I.,
Hartig, G.~F., Ford, H.~C., Dopita, M.~A., Kriss, G.~A.,
Pei, Y.~C., Dressel, L.~L., \& Harms, R.~J.\ 1998, ApJ, 493, L83

\bibitem[Vaceli et al.~1997]{vaceli}
Vaceli, M.~S., Viegas, S.~M., Gruenwald, R., \& de Souza, 
R.~E.~1997, AJ, 114, 1345

\bibitem[Veron-Cetty \& Veron 1993]{veron}
Veron-Cetty, M.-P., \& Veron, P.\ 1993,  ESO Sci. Rep., 13, 1

\bibitem[Zeldovich \& Novikov 1964]{zeldovich}
Zeldovich, Ya.~B., \& Novikov, I.~D.~1964, 
Dokl.~Akad.~Nauk SSSR, 155, 1033

\bibitem[Clavel, Wamsteker \& Glass 1989]{clavel3}
\bibitem[Ho et al.~(1995b)]{hob_}


\end{thebibliography}
\end{document}